\documentclass[a4paper]{article}
\pdfoutput=1

\usepackage{INTERSPEECH2020}

\title{An Acoustic Segment Model Based Segment Unit Selection Approach to Acoustic Scene Classification with Partial Utterances}

\name{Hu Hu$^{1}$, Sabato Marco Siniscalchi$^{2,1}$, Yannan Wang$^{3}$, Xue Bai$^{4}$, Jun Du $^{4}$, Chin-Hui Lee$^{1}$ }
\address{$^1$School of Electrical and Computer Engineering, Georgia Institute of Technology, USA \\
$^2$Computer Engineering School, University of Enna, Italy \\
$^3$Tencent Media Lab, Tencent Corporation, Shenzhen, Guangdong, China\\
$^4$University of Science and Technology of China, HeFei, China}
\email{huhu@gatech.edu, marco.sinsalchi@unikore.it, yannanwang@tencent.com, byxue@mail.ustc.edu.cn, jundu@ustc.edu.cn, chl@ece.gatech.edu}

\begin{document}

\maketitle
\begin{abstract}

In this paper, we propose a sub-utterance unit selection framework to remove acoustic segments in audio recordings that carry little information for acoustic scene classification (ASC). 
Our approach is built upon a universal set of acoustic segment units covering the overall acoustic scene space.
First, those units are modeled with acoustic segment models (ASMs) used to tokenize acoustic scene utterances into sequences of acoustic segment units. Next, paralleling the idea of stop words in information retrieval, stop ASMs are automatically detected. Finally, acoustic segments associated with the stop ASMs are blocked, because of their low indexing power in retrieval of most acoustic scenes. In contrast to building scene models with whole utterances, the ASM-removed sub-utterances, i.e., acoustic utterances without stop acoustic segments, are then used as inputs to the AlexNet-L back-end for final classification. On the DCASE 2018 dataset, scene classification accuracy increases from 68\%, with whole utterances, to 72.1\%, with segment selection. This represents a competitive accuracy without any data augmentation, and/or ensemble strategy. Moreover, our approach compares favourably to AlexNet-L with  attention.
\end{abstract}


\section{Introduction}
The aim of the acoustic scene classification (ASC) refers to the task of identifying real-life sounds into environment classes, such as metro station, street traffic, or public square. An acoustic scene sound contains much information and rich content, which makes accurate scene prediction difficult. ASC has been an attracting research field for decades, and the IEEE Detection and Classification of Acoustic Scenes and Events (DCASE) challenge \cite{dcase2016, dcase2017, dcase2018} provides the benchmark data and a competitive platform to promote sound scene research and analyses. In recent years, we have witnessed that the deep neural networks (DNNs) have gradually dominated the design of top ASC systems, and the main ingredient of their success is the application of deep convolutional neural networks (CNNs) \cite{asc-cnn2, asc-cnn3, asc-cnn4, dcase-2020-rank2}. Furthermore, with the use of advanced deep learning techniques, such as attention mechanism \cite{asc-attention1, asc-attention2, asc-attention4} and deep network based data augmentation  \cite{asc-mixup1, asc-gan1, asc-aug1}, a further boost in ASC system performances can be obtained.

In this study, we leverage upon acoustic segment models (ASMs) as an indicator of the indexing power of the input audio segment units with respect to the acoustic scenes being classified. A set of ASM models is employed to carry out acoustic segment selection in the front-end. An initial ASM sequence for each given audio recording is obtained by unsupervised segmentation and clustering. Next, we use Gaussian mixture model (GMM)- or deep neural network (DNN)- hidden Markov model (GMM/DNN-HMM) \cite{gmmhmm, dnnhmm} to model the ASM sequences in a semi-supervised manner. Thus, each audio recording is segmented (tokenized) into a sequence of acoustic segment units, each having its relative ASM unit index. In this work, the terms acoustic segment units, acoustic segments, sub-utterance units, and tokens, are used interchangeably.
A similar strategy to detect the stop words in information retrieval \cite{stopwords0} is adapted, and a set of stop ASMs is identified using the training data. Stop ASMs represent meaningless ASM units, which carrying very low indexing power in retrieving most acoustic scenes. Just like word 'the', 'an' and 'or' in document retrieval problems, stop ASMs are therefore not useful to identify the target scene class. Those stop ASMs are used at a front-end level to block all of audio segments consisting of sequences of acoustic frames, belonging to those stop ASM models in both the training and evaluation stages. In doing so, noisy acoustic segments are eliminated in building models at the training stage, and not sent to the back-end acoustic scene classifier during the classification stages. The proposed approach is evaluated on DCASE 2018 Task1a data set. Our experiments demonstrate that our solution improves an AlexNet like system, dubbed AlexNet-L, boosting the classification accuracy from 68.0\% to 72.1\%. The latter is a competitive result since neither data augmentation nor system combination are used. Furthermore, our segment-selection scheme with ASMs compares favourably with a recently proposed CNN classifiers using an ASM-based attention mechanism \cite{ustc2}.


\section{Related Work and Our Contributions}
\label{sec:related}
Several approaches based on feature learning have been proposed for the ASC task. For example, low-level features \cite{asc-lowfeat1, asc-lowfeat2, asc-lowfeat3}, which are directly extracted from the input signal at the front-end level, are thoroughly investigated to boost ASC performance. With deep models, mid-level features \cite{asc-ivector, asc-xvector, asc-midfeat1, openl3} are instead induced from a DNN hidden layer, which takes into account the overall information embedded in training set. In addition to these low-level or mid-level features, the use of raw waveform to feed end-to-end systems has also been investigated in \cite{asc-rawwave1, asc-rawwave2}. However, to the best of the authors' knowledge, all of those feature learning works on ASC employ the whole input audio recording at the input layer to obtain high-dimensional feature vectors, and there aren't investigations concerned with segment selection at a front-end level.

From a  human listening perspective, sound recognition is often guided by detecting prominent acoustic events and/or audio cues useful to identify particular acoustic scenes \cite{asc-human}. For example, human listeners may leverage upon a car horn sound to determine that it is from a street traffic scene, or a loud plane engine sound to determine it is from an airport. Those sounds generated from car horns and plane engines, have stronger indexing power than other sounds for classifying these two acoustic scenes. Hence we argue that we are bound to get better ASC accuracy if we can block acoustic segments with little indexing power. Our idea could be related to an attention mechanism \cite{asc-attention1, asc-attention2, asc-attention4, ustc2}, which uses an ad-hoc internal connectionist block and a huge amount of data to weight hidden internal representations accordingly to its salience to the target outputs.
However, an attention mechanism requires extra amount of parameters, and the performance highly depends on the model tuning. Our approach introduces a well-known approach from the information retrieval field \cite{stopwords0} to detect meaningless sound events, and the experimental evidence confirms our claim.

In order to find the semantic salience of sound events, we use ASMs. ASMs are a set of self-organized sound units that are intended to cover the overall acoustic characteristics using available training data \cite{asm1}. The ASM framework has recently been adopted in many audio sequence classification tasks, such as language identification \cite{asm-lid}, speaker identification \cite{asm-si}, emotion recognition \cite{asm-emotion}, music genre classification \cite{asm-music} and the acoustic scene classification \cite{ustc1}. As for ASC, it makes the assumption that the acoustic characteristics of all scenes can be covered by a universal set of acoustic units. Thus, input audio recordings can be transformed into ASM sequences, which are in turn processed by latent semantic analysis (LSA) \cite{lsa} to obtain  feature vectors with semantic information. Finally, a CNN based ASC system with an attention mechanism using ASM units is proposed in \cite{ustc2}. Different from the conventional ASM framework, in this work, ASM sequences are not used for a follow-up feature extraction process. In the experimental section, we demonstrated that our front-end solution outperforms that with the attention mechanism in \cite{ustc2}.

\section{Acoustic Segment Modeling}
\label{sec:asm}
Like the phoneme representation for the speech utterance, we assume that the sound characteristics of acoustic scenes can also be covered by a universal set of acoustic units. The ASM approach aims to build a tokenizer to transfer the scene audio into a sequence of ASMs, i.e., the acoustic units specified in an acoustic inventory. The ASM sequence is generated in two main steps: (i)  an unsupervised approach is used to seed the initial ASMs, with each acoustic unit having a fixed length (acoustic segment), (ii) either a GMM-HMM or DNN-HMM system is built on top of the initial ASMs and then used to generate the ASM sequence for a given audio recording.

\begin{figure}[t]
  \vspace{1mm}
  \centering
  \includegraphics[width=0.9\linewidth]{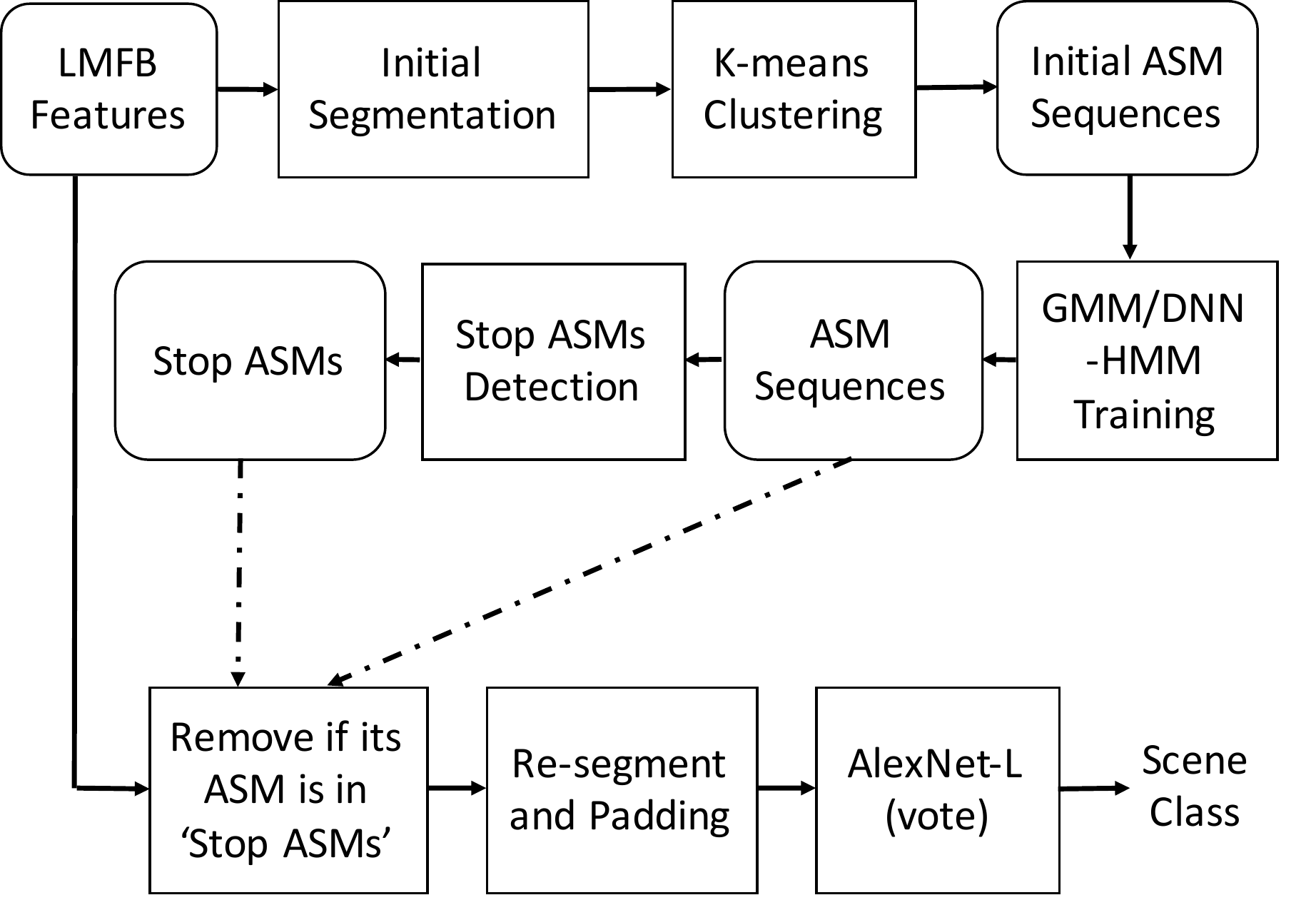}
  \caption{The framework of proposed ASM-guided segment selection approach for ASC.}
  \label{fig:framework}
  \vspace{1mm}
\end{figure}

\subsection{Initial ASM Sequence Generation}


ASM initialization is a critical factor for the success of the ASM framework. The initial ASM sequence generation is performed at a feature level, i.e. log-mel filter bank (LMFB) energies, or mel frequency cepstral coefficients (MFCCs). First, a given input audio recording is divided into a sequence of fixed-length segments. In our experiments, an audio recording is split into $50$ acoustic segments, each having  $20$ LMFB or MFCC frames. The arithmetic mean of these $20$ frames is used to generate a single feature vector representing the whole segment. Next, all generated feature vectors for the training material are used with the K-means clustering algorithm to find a set of centroids. Audio segments are grouped into a small number of acoustic classes (each class is an ASM), to represent the whole acoustic space scattered by the training data. We do not leverage any prior knowledge when building our acoustic inventory; therefore, the set of ASMs and its corresponding model arises in an unsupervised manner. Finally, the centroids can be used to map a given audio recording into an initial ASM sequence. Each initial ASM sequence has a fixed number of $50$ ASM units, since we have split the input recording into $50$ fixed-length segments.

\subsection{GMM/DNN-HMM Training}
The initial ASMs can provide a rough segmentation of the audio scene, where segments have fixed-length in terms of the number of audio frames, which does not adhere to real scenarios. A finer segmentation result can be obtained leveraging the HMM framework, in which the state probability density functions (pdfs) can be obtained with a GMM or DNN. We therefore deployed both GMM-HMM and DNN-HMM tokenizers as follows:
\begin{enumerate}
    \item Seed a GMM-HMM for each ASM unit based on the initial ASM segmentation.
    \item Use the model obtained in Step 1 and perform Viterbi decoding on all the training utterances.
    \item Train the model with the new transcriptions generated in Step 2.
    \item Repeat Step 2 and Step 3 until convergence.
    \item Build a DNN-HMM using the GMM-HMM and new ASM segmentation.
\end{enumerate}


\section{Front-end Segment Selection via ASM}
\label{sec:process}
For acoustic scenes, differently from spoken utterances, only few real meaningful segments characterize the whole scene. For example, a car horn sound is an import sign to determine  a street traffic scene, but many other segments in that acoustic scene do not carry any key information to make the correct classification. However, it's not easy to detect meaningful segments for most of the acoustic scenes. Therefore, more and more ASC systems simply use a CNN end-to-end approach to learn the mapping from the input audio recordings to the output scene class . Nonetheless, CNNs are good at extracting local features but not for overall segment selection.  ASMs are used in our work to find useful acoustic segments, which have high indexing power with respect to the target acoustic scene. If useless segments can be removed in the front-end processing stage, that will be beneficial to the back-end classifier, as proven in the experimental section.

The proposed ASM-based front-end segment selection framework for sub-utterance acoustic scene classification is shown in Figure~\ref{fig:framework}. The dashed lines indicate where ASM sequences and stop ASMs are used. Stop ASMs, and ASM sequences are generated in the segment selection block to remove consecutive feature frames not useful for final scene classification. The stop rules and segment selection steps are described in detail in the following section.

\subsection{Stop ASMs Detection}
\label{sec:stopASM}
Stop ASMs, as the name reveals, takes inspiration from stop words in information retrieval \cite{stopwords0}. Given the inventory of ASMs, stop ASMs are a subset of the original ASMs that does not carry much information for retrieving the target acoustic scenes. Compared with other ASMs in the inventory, stop ASMs have either a lower indexing power, or no indexing power at all when it comes to make the final classification decision. We use $D$ to denote the total number of ASMs in the inventory, and $N$ to denote the total number of utterances in the training set. In this study, we consider four different methods to detect the stop ASMs \cite{stopwords1, stopwords2, stopwords3}:
\begin{itemize}
    \item Mean Probability (MP): it is the average probability of each unit $M_j$ in ASM sequences of the training set. MP considers the frequency of each ASM and is calculated by
    \begin{equation}
      MP (M_j) = \frac{\sum_{i=1}^N P_{i,j}}{N},
    \end{equation}
    in which $P_{i,j}$ is the probability of the ASM unit $M_j$ in utterance $U_i$, and calculated by dividing its frequency by the total number of the ASMs in $U_i$.

    \item Inverse Document Frequency (IDF): it measures how much information the ASM provides to reflect the importance of each ASM. IDF is calculated by
    \begin{equation}
        IDF (M_j) = \log \frac{N + 1}{N_j + 1} ,
    \end{equation}
    in which $N_j$ is total number of times the ASM unit $M_j$ appears in the training utterances.

    \item Variance of Probability (VP): it considers the variance of each ASM unit. VP is calculated by
    \begin{equation}
        VP (M_j) = \frac{\sum_{i=1}^N (P_{i,j} - MP(M_j))^2}{N}.
    \end{equation}

    \item Statistical Values (SATs): this metric considers both the mean and the variance. If an ASM unit has high SAT values, it implies that $M_j$ occurs frequently and uniformly in all the training utterances, and $M_j$ is very likely to be a stop ASM. SAT is calculated by
    \begin{equation}
        SAT (M_j) = \frac{MP(M_j)}{VP(M_j)^{1/2}} .
    \end{equation}

\end{itemize}

In our experiments, we select the top $P$ ASMs for functioning as stop ASMs. The segments whose ASMs are in stop ASMs dominate in most utterances.

\subsection{Data Processing with Stop ASMs}
The front-end  processing is performed with stop ASMs and ASM sequences. As shown in Figure~\ref{fig:framework}, for a given input with an ASM sequence, the corresponding LMFB feature frames will be removed if their ASMs are stop ASMs. The remaining feature fragments will be re-segmented and padded into a group of fixed-length acoustic segments. In our experiments, if a fragment is more than $20$ frames, we will divide it into segments with the length of $20$ frames, otherwise, we will perform zero-padding to make it having $20$ frames. Hence, each acoustic scene would eventually have a different number of segments, which depends on the number of frames that have been blocked at the front-end level. After re-segmenting and padding process, the new generated fixed-length segments are fed into the back-end classifier. Each segment is assigned to a scene class by AlexNet-L back-end classifier, and the final scene class is obtained via majority voting among all segments.


\section{Experiments and Result Analysis}
\label{sec:exp}
\subsection{Experimental Setup}
The proposed approach is evaluated on the DCASE 2018 Task1a development data set \cite{dcase2018}. It contains $24$ hours of acoustic scene audio recorded with the same device at a $48$kHz sampling rate in $10$ different acoustic scenes. Following the official recommendation, the development data set is divided into training and test sets containing 6122 and 2518 utterances, respectively. For each 10-second binaural signal, STFT with $2048$ points is applied separately on the left and right channels, with a window length of $25$ms and an overlap length of $10$ms. Mel filter-banks  with $128$ bins are applied to obtain the log-mel filter bank (LMFB) features. Our ASC baseline system is based on the AlexNet \cite{alexnet} model. Nonetheless, different from the original AlexNet, we reduce the parameter size due to internal resource constrains. The baseline is denoted as AlexNet-L in the rest of this work. It has five convolutional layers with a kernel size of $4 \times 4$, and two fully connected layers with hidden dimension of $1024$. Each layer consists of convolution, batch normalization, ReLU activation function and max pooling.
AlexNet-L is trained with a stochastic gradient descent (SGD) algorithm with a cosine based learning rate scheduler.  Each input utterance is segmented to $20$ frames ($0.2$ seconds per segment). After AlexNet-L classification, the final scene class of the input waveform is voted by a majority using the classification result on each segments.

In the acoustic segment modeling stage, the initial segment length is set to 20 frames, which is the same with the segment length in the baseline AlexNet-L. Hence each utterance is divided into $50$ segments. The size $D$ of the ASM inventory is set to 64. According to our experiments, $D$ is robust to the parameter setting.  GMM-HMMs are powerful for modeling sequential data, and we here to refine the initial tokenization phase. A left-to-right HMM topology is used in each 6-state GMM-HMM. In DNN-HMM, the DNN has six hidden layers, each having 2048 neurons. The output layer estimates the state probability density function (pdf) of the 64 HMMs. During the stop ASMs detection, top-3 ASMs function as stop ASMs for each metric discussed in Section \ref{sec:stopASM}. After removing segments whose ASMs are in the stop ASMs set, the remaining acoustic fragments are re-segmented and padded to obtain $20$ acoustic frames per segment. As done with AlexNet-L, majority voting is used to decide the final scene class.

\subsection{Stop ASMs Detection Results}
The stop ASMs detection result is shown in  Table~\ref{tab:stopasms}. Different detection criteria lead to a  different set of stop ASMs. We selected the top three ASMs according to highest MP, lowest IDF, lowest VP and highest SAT, as our stop ASMs. From Table~\ref{tab:stopasms}, we can notice that some ASMs, such as $M_8$, $M_{14}$, appear independently of the metrics. The latter implies that some acoustic segments do satisfy our assumption, that is, there are acoustic segments having low or no indexing power for scene classification.  Stop ASMs found by GMM-HMM and DNN-HMM are similar. When using the SAT criterion, the same set of stop ASMs is found independently of the tokenizer. However, although different metrics can select the same stop ASMs,  the technique to obtain ASMs  would lead to different segment boundaries, which eventually affects final classification results.

\begin{table}[ht]
\centering
\caption{Stop ASMs detection results with different metrics. $M_i$ indicates $i$-th ASM in the ASM inventory. Initial ASMs imply that initial ASM sequences are used in stop ASMs detection. GMM-HMM and DNN-HMM refer to sequences obtained with the those models.}
{\footnotesize
\begin{tabular}{|l|c|c|c|}
\hline
Metric & Initial ASMs & GMM-HMM       & DNN-HMM       \\
\hline
MP  & $M_8,M_{35},M_{43}$& $M_{21},M_{35},M_{43}$ &  $M_8,M_{35},M_{43}$\\
\hline
IDF & $M_8,M_{10},M_{14}$ &  $M_8,M_{10},M_{51}$  &  $M_8,M_{10},M_{14}$ \\
\hline
VP  & $M_4,M_{39},M_{53}$ &  $M_{39},M_{42},M_{56}$ &  $M_{39},M_{55},M_{56}$ \\
\hline
SAT &$M_8,M_{14},M_{52}$&  $M_8,M_{14},M_{52}$  &  $M_8,M_{14},M_{52}$ \\
\hline
\end{tabular}
\vspace{-2mm}
}

\label{tab:stopasms}
\end{table}

\subsection{ASC Experimental Results}
The proposed ASC system is shown in Figure~\ref{fig:framework}, and related  experimental results are given in Table~\ref{tab:res}. For a comprehensive evaluation, the high-resolution attention network with ASM (HRAN-ASM) system proposed in \cite{ustc2} is also implemented, and its classification results are reported. In particular, we have adopted the two attention modules with ASM embedding into our AlexNet-L baseline model. The first two rows in Table \ref{tab:res} are official baseline \cite{dcase2018} and our AlexNet-L baseline. The AlexNet-L attains a classification accuracy of $68.0\%$, which is improved to $69.5\%$ by with the HRAN-ASM in the third row. Although our experiments are conducted with the same data sets adopted in \cite{ustc2}, experimental results are slightly different because of different speech features and models used in our work.

\begin{table}[t]
\centering
\caption{Evaluation results on DCASE 2018 Task1a data set.}
\begin{tabular}{l||c}
\hline
\hline
Model               & Accuracy \% \\
\hline
\hline
 Official baseline \cite{dcase2018} & 59.7 \\
 AlexNet-L  baseline & 68.0        \\
\ \ + HRAN-ASM \cite{ustc2}     & 69.5        \\
\hline
\ \ + initial ASM (SAT) & 70.1 \\
\ \ + ASM-GMM-HMM (SAT) & 71.6        \\
\ \ + ASM-DNN-HMM (SAT) & 72.1    \\
\hline
\hline
\end{tabular}
\label{tab:res}
\vspace{-2mm}
\end{table}

Table~\ref{tab:res} lists experimental results obtained with the proposed approach in the last three rows when SAT is used as the metric to extract the stop ASMs. We can use our acoustic segment blocking approach with three different ASM tokenizers, namely (i) the initial unsupervised ASM (initial ASM), (ii) GMM-HMM, and (iii) DNN-HMM. In case (i), stop ASMs are first detected using the tokenization obtained with initial ASM.
Since the segment length of initial ASM sequences and the input sequences to AlexNet-L are the same, we can directly block the whole segment of the input sequence if its corresponding ASM token is in the set of stop ASMs. Thus, the blocking operation is based on the segment level, in which the re-segmenting and padding is not needed on processed segments.
Although initial ASMs are simple models, AlexNet-L accuracy can be boosted from $68.0\%$ to $70.1\%$ (compare first and third rows). In cases (ii) and (iii) with either GMM-HMM or DNN-HMM tokenizers, we can obtain more precise ASM sequences, which in turn improve the final ASC accuracy as shown in the last two rows of Table~\ref{tab:res}. In details, GMM-HMM boosts AlexNet-L classification up to $71.6\%$.  DNN-HMM  can deliver more accurate alignments and boundaries for each ASM segment, which leads to a final scene classification accuracy of $72.1\%$, which representing a $4.1\%$ absolute improvement when compared to AlexNet-L.

The above results allow us to conclude that: (a) the ASM-based segment selection approach  can significantly improve ASC system, attaining a final classification accuracy of 72.1\%, which is a competitive performance given that data augmentation and ensemble methods have not been used in our work, and (b) the proposed solution outperforms  AlexNet-L with the attention mechanism initialised with ASM, which the later is reported to compare favourably against self-attention  in \cite{ustc2}. The latter demonstrates that our front-end segment selection approach outperforms a more standard attention scheme.

\subsection{Evaluation with Different Detection Metrics}
The effect of different stop ASMs metrics is shown in Table~\ref{tab:res-metric}. The initial ASM sequences are used to evaluate different metrics. From Table \ref{tab:res-metric},  we can see that different stop ASMs detection metrics result in different classification outcomes. Using MP with initial ASM sequences does not lead to any improvement over AlexNet-L. SAT shows the best performance among all metrics. Those results makes sense since  SAT considers both the mean and the variance of the distribution of each ASM unit.

\begin{table}[ht]
\centering
\caption{Results with different stop ASMs detection metrics using initial ASMs.}
\begin{tabular}{l||c}
\hline
\hline
Model               & Accuracy \% \\
\hline
\hline
AlexNet-L baseline & 68.0        \\
\hline
\ \ + initial ASM (MP) & 68.0 \\
\ \ + initial ASM (IDF) & 68.7 \\
\ \ + initial ASM (VP) & 69.3 \\
\ \ + initial ASM (SAT) & 70.1 \\
\hline
\hline
\end{tabular}
\label{tab:res-metric}
\vspace{-1mm}
\end{table}

\section{Summary}
\label{sec:con}
In this paper, instead of using whole utterances for scene modeling, we propose an ASM based front-end segment selection approach to acoustic scene classification. The overall framework is based on two modules: (i) acoustic segment modeling and selection, and (ii) CNN based classification. ASMs are first generated in an unsupervised manner and refined with GMM/DNN-HMM models. Then stop ASMs detection is performed using ASM sequences for training data.  ASM sequences and stop ASMs are used for segment selection before CNN classifier. It implies that segments with low or no indexing power are removed. The proposed approach is evaluated on DCASE 2018 Task1a, and experimental evidences demonstrate the viability of sub-utterance ASC. A classification accuracy of $72.1\%$ is obtained, which  is highly competitive for single system and no data expansion. 

\clearpage
\bibliographystyle{IEEEtran}

\bibliography{mybib}

\end{document}